\documentclass[12pt]{article} % single column, double spaced

\usepackage{osajnl}
\usepackage{amsmath}
\usepackage{graphicx}

\newcommand{\tr}{\operatorname{tr}}
\newcommand{\re}{\operatorname{Re}}
\newcommand{\diag}{\operatorname{diag}}

\begin{document}

%\twocolumn[ %% activate for two-column option

\title{A three-dimensional degree of polarization based on Rayleigh scattering}

\author{Mark R Dennis}

\address{School of Mathematics, University of Southampton, Highfield, Southampton SO17 1BJ, UK}

\begin{abstract}
A measure of the degree of polarization for the three-dimensional polarization matrix (coherence matrix) of an electromagnetic field is proposed, based on Rayleigh scattering.
The degree of polarization, due to dipole scattering of the three-dimensional state of polarization, is averaged over all scattering directions.
This gives a well-defined purity measure, which, unlike other proposed measures of the three-dimensional degree of polarization, is not a unitary invariant of the matrix.
This is demonstrated and discussed for several examples, including a partially polarized transverse beam.
\end{abstract} 


% ] %% activate for two-column option

\section{Introduction}\label{sec:int}

Polarization is a fundamental property of electromagnetic fields.
Its effects are particularly important when the field's 3-dimensional, vectorial character plays a role, such as radiation from sources, the optical near-field, tightly-focussed beams, and scattering.
In these situations, the conventional theory of partial paraxial polarization fails.
There has been recent interest in extending the theory of partial polarization to these nonparaxial situations, particularly the notion of degree of polarization (e.g. Refs. \onlinecite{brosseau:fundamentals,sskf:degree,dennis:geometric,edpw:degree,luis:degree,lskf:degree,ed:degree} and references therein).

The transverse polarization properties of optical beams are well understood.\cite{brosseau:fundamentals,wolf:coherence}
The theory is based on the hermitian $2\times 2$ polarization matrix $\boldsymbol{\rho}_2,$ defined 
\begin{equation}
   \boldsymbol{\rho}_2 = \frac{\langle \boldsymbol{E}^{\ast}\otimes\boldsymbol{E}\rangle}{\langle |\boldsymbol{E}|^2 \rangle},\qquad 
   \rho_{2,ij} = \frac{\langle E_i^{\ast} E_j \rangle}{\langle | \boldsymbol{E} |^2 \rangle},
   \label{eq:edef}
\end{equation}
with averaging over some ensemble of transverse field vectors $\boldsymbol{E} = (E_1,E_2).$
This matrix has a unique linear decomposition: $\boldsymbol{\rho}_2 = P_2 \boldsymbol{\rho}_{\mathrm{pure}} + (1-P_2) \boldsymbol{\rho}_{\mathrm{2,un}},$ where $\boldsymbol{\rho}_{\mathrm{pure}}$ is the polarization matrix (single-point coherence matrix) of a pure polarization state, $\boldsymbol{\rho}_{\mathrm{2,un}}$ is the trace-normalized $2\times 2$ identity matrix (completely unpolarized), and $0 \le P_2 \le 1.$
$P_2$ is the modulus of the difference between the eigenvalues of $\boldsymbol{\rho}_2,$ so
\begin{equation}
   P_2^2 = \frac{2\tr \boldsymbol{\rho}_2^2 - (\tr \boldsymbol{\rho}_2)^2}{(\tr \boldsymbol{\rho}_2)^2}.
   \label{eq:p2def}
\end{equation}
When $\boldsymbol{\rho}_2$ is decomposed via the basis of Pauli matrices, the coefficients are the Stokes parameters $S_1, S_2, S_3$ ($S_0 = 1$ since $\tr\boldsymbol{\rho}_2$ is normalized).
The sum of squares $S_1^2 + S_2^2 + S_3^2  = P_2^2;$ on propagation through rotators and retarders (represented by unitary transformations on $\boldsymbol{\rho}_2$), the Stokes parameters may change, but $P_2$ does not.

The theory of 3-dimensional partially coherent fields is less well developed.
Even in fully polarized fields, the Stokes parameter description fails since the plane of the polarization ellipses varies with position.
Furthermore, in partially polarized fields, polarization states with ellipses in different planes may be incoherently mixed.
Several measures of the 3-dimensional degree of polarization have been proposed and discussed\cite{sskf:degree,edpw:degree,luis:degree,lskf:degree}; these measures, discussed below, have the property of unitary invariance: they only depend on the eigenvalues of the 3-dimensional polarization matrix $\boldsymbol{\rho}_3,$ by analogy with unitary invariance of $P_2.$
In terms of a measure on the eigenvalues, $P_2$ is clearly unique (up to monotonic transformations, such as squaring), since after trace normalization, the two eigenvalues have only one freedom (their difference).
However, for $3 \times 3$ polarization matrices, there are two such freedoms, and there is no unique mathematical definition for the degree of polarization; it is therefore appropriate to examine measures that emerge naturally out of physical processes, even if desirable mathematical requirements have to be relaxed.
Clearly, one such requirement that cannot be satisfied in three dimensions is the decomposition of the polarization matrix into purely polarized and purely unpolarized parts;\cite{edpw:degree} the intermediate eigenvalue must also play a role in any matrix decomposition (the largest and smallest eigenvalues corresponding to the completely polarized and unpolarized parts).

This is the optical analog of a well-known problem in quantum mechanics, namely the non-uniqueness of the decomposition of a general density matrix.\cite{peres:quantum,sakurai:modern,fraassen:quantum}
Thus, the definition of the degree of polarization is chosen here to correspond to that of a purity measure in quantum mechanics: a purely polarized state (with eigenvalues $1,0,0$) always has measure 1, and the completely unpolarized state ($1/3$ times the identity matrix), isotropic in all directions, has measure 0; any other state of polarization has a value in between.
For $\boldsymbol{\rho}_2,$ $P_2$ clearly satisfies this, and is effectively unique.
However, under this definition, the quantity proposed in Ref. \onlinecite{edpw:degree} (the difference between the two larger eigenvalues of $\boldsymbol{\rho}_3$) -- motivated by the eigenvalue decomposition of polarization -- is not a purity measure, as anisotropic states of 3-dimensional polarization, which have directional information, are counted as completely unpolarized (for instance, a completely unpolarized paraxial beam has directional information (its propagation direction), but its eigenvalues are $1,1,0$).
In quantum mechanics and paraxial optics, such unitary invariance is a physical requirement, but there seems to be no strong argument for this in the nonparaxial case,\cite{dennis:geometric} a point discussed towards the end of this article.

Here, I describe an alternative measure for the degree of polarization in three dimensions, based on isotropic Rayleigh scattering, which is not a unitary invariant of $\boldsymbol{\rho}_3.$
This is defined and discussed in the following section, then in Section \ref{sec:exs}, is compared with various other 3-dimensional polarization measures for various specific examples of $\boldsymbol{\rho}_3.$
The final section consists of a discussion of various issues associated with the Rayleigh-defined degree of polarization and non-unitarity.

\section{The Rayleigh-defined degree of polarization}\label{sec:def}

Rayleigh scattering is a fundamental 3-dimensional optical phenomenon, and is of primary importance in scattering theory.\cite{newton:scattering,jackson:classical,chandrasekhar:radiative}
At an isotropic scatterer, of dimension much smaller than the optical wavelength, the possibly incoherent electromagnetic field is represented by density matrix $\boldsymbol{\rho}_3$ (defined analogously to $\boldsymbol{\rho}_2$ in Eq. (\ref{eq:edef})).
Partially polarized rays are scattered in all directions $\theta, \phi;$ the properties of the scattered ray in $\theta, \phi$ are determined by the polarization matrix\cite{newton:scattering, jackson:classical}
\begin{equation}
   \boldsymbol{\rho}_2(\theta,\phi) = \frac{3\sigma}{8\pi} \mathbf{p}_{\bot} (\theta,\phi) \cdot \boldsymbol{\rho}_3 \cdot \mathbf{p}_{\bot} (\theta,\phi),
   \label{eq:proj}
\end{equation}
where $\sigma$ is the total scattering cross-section, and $\mathbf{p}_{\bot}(\theta, \phi)$ is a projection matrix, projecting into the plane perpendicular to the unit vector $\boldsymbol{u}(\theta,\phi)$ in the direction $\theta, \phi.$
Although $\boldsymbol{\rho}_2(\theta,\phi)$ is $3\times 3,$ it is null in (appears transverse to) the direction $\boldsymbol{u}(\theta,\phi)$ by definition.
$\boldsymbol{\rho}_2(\theta,\phi)$ is not trace-normalized, its trace being the intensity $I(\theta,\phi)$ scattered in the $\theta,\phi$-direction:
\begin{equation}
   I(\theta, \phi) = \tr\boldsymbol{\rho}_2(\theta,\phi)
   \label{eq:itp}
\end{equation}
(this is the differential scattering cross section\cite{newton:scattering, jackson:classical}).
The scattering mechanism is here idealized such that the scatterer experiences no recoil (there is no center-of-mass motion in the scattering). 

The degree of polarization of the scattered light depends on the scattering direction, but is otherwise similar to Eq. (\ref{eq:p2def}),
\begin{equation}
   P_2^2(\theta,\phi) = \frac{2\tr \boldsymbol{\rho}_2^2(\theta,\phi) - [\tr \boldsymbol{\rho}_2(\theta,\phi)]^2}{[\tr \boldsymbol{\rho}_2(\theta,\phi)]^2}
   \label{eq:p2tp}
\end{equation}
where the denominator equals $I(\theta, \phi).$
The {\em Rayleigh-defined degree of polarization} $P_{\mathrm{Ray}}$ is taken to be the average over all directions of $P(\theta,\phi),$ weighted with respect to $I(\theta, \phi):$
\begin{equation}
   P_{\mathrm{Ray}} = \frac{\int_0^{2\pi} d \phi \int_0^{\pi} d \theta \, \sin \theta \, P_2(\theta,\phi) I(\theta, \phi)}{\int_0^{2\pi} d \phi \int_0^{\pi} d \theta \, \sin \theta \, I(\theta, \phi)}
   \label{eq:pray}
\end{equation}
where the denominator is the total scattering cross-section $\sigma.$

This process can be visualized using the geometric interpretation of the 3- dimensional polarization matrix\cite{dennis:geometric}.
Choosing appropriate cartesian axes, $\boldsymbol{\rho}_3$ can be written
\begin{equation}
   \boldsymbol{\rho}_3 = \left(\begin{array}{ccc}
M_1 & -i N_3 & i N_2 \\
i N_3 & M_2 & -i N_1 \\
-i N_2 & i N_1 & M_3 \end{array} \right),
   \label{eq:rho3def}
\end{equation}
where $M_j, N_j$ are real, and $1 \ge M_1 \ge M_2 \ge M_3 \ge 0.$
The real part, $\mathbf{M} = \diag\{M_1,M_2,M_3\},$ is interpreted as the moment of inertia ellipsoid of the polarization ellipse ensemble.
The imaginary part is an axial vector, $\boldsymbol{N} = (N_1, N_2, N_3),$ the average angular momentum vector of the ensemble.
A completely unpolarized ensemble (representing, for instance, black-body radiation) is therefore represented by a sphere, and a pure polarization state by an ellipse with orthogonal vector (whose length is the ellipse area).
For a pure polarization state, the 1 (2)-axis is the major (minor) axis of the polarization ellipse, and the 3-axis is its angular momentum direction: the flat ellipsoid ($M_3 = 0$) has axes proportional to the polarization ellipse axes squared ($M_j  = |E_j|^2/|\boldsymbol{E}|^2, j = 1,2$) and $|\boldsymbol{N}| = |N_3| = |\boldsymbol{E}^{\ast} \times \boldsymbol{E}|/2 = \sqrt{M_1 M_2}.$
The ellipsoid and vector can change under unitary transformation.

\begin{figure}
   \centerline{\includegraphics[width=6cm]{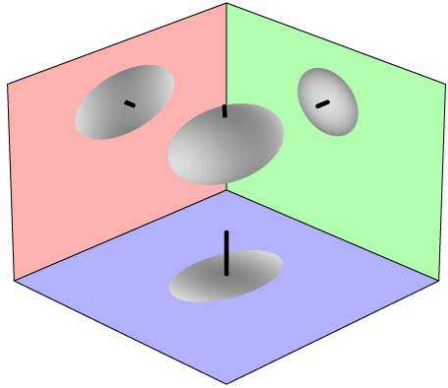}}
   \caption{Geometric representation of polarization matrix $\boldsymbol{\rho}_3$ as ellipsoid and vector. 
   The ellipses and vectors for $\boldsymbol{\rho}_2(\theta,\phi)$ in three orthogonal projection directions are also shown.
   }
   \label{fig:geb}
\end{figure}

The ellipsoid $\mathbf{M}$ and vector $\boldsymbol{N}$ generalize features of $\boldsymbol{\rho}_2;$ the real part $\re \boldsymbol{\rho}_2$ defines an ellipse, and the imaginary part, proportional to the Stokes parameter $S_3,$ is the overall angular momentum perpendicular to the transverse plane.
The polarization state $\boldsymbol{\rho}_2(\theta,\phi)$ of a scattered ray is geometrically determined by projection; the transverse ellipse of its real part is the projection of the ellipsoid in the plane perpendicular to $\theta, \phi,$ and its imaginary part is the projection of $\boldsymbol{N}$ into $\boldsymbol{u}(\theta,\phi),$ as depicted in Fig. \ref{fig:geb} projecting in the coordinate axis directions.
$P_{\mathrm{Ray}}$ is the average of the degree of polarization over all these projections.

Although $P_{\mathrm{Ray}}$ depends only on the polarization matrix $\boldsymbol{\rho}_3,$ it cannot be simply expressed in terms of the matrix elements.
Nevertheless, it is a purity measure on the density matrix $\boldsymbol{\rho}_3,$ as defined above.
If $\boldsymbol{\rho}_3$ represents a pure state, its eigenvalues are $1,0,0,$ and so $P_2(\theta,\phi)$ (the difference divided by the sum of eigenvalues of $\boldsymbol{\rho}_2(\theta,\phi)$) is 1 for almost all directions, thus $P_{\mathrm{Ray}} = 1.$
Conversely, $P_{\mathrm{Ray}} = 1$ in Eq. (\ref{eq:pray}) only if $P(\theta,\phi) = 1$ for almost all directions, which only occurs if one eigenvalue of $\boldsymbol{\rho}_2(\theta,\phi)$ is zero, implying $\boldsymbol{\rho}_3$ is pure.
Similarly, $P_{\mathrm{Ray}} = 0$ if and only if $\boldsymbol{\rho}_3$ is the completely unpolarized 3-dimensional matrix $\boldsymbol{1}/3.$

In this argument, `almost all' assumes its technical sense, that is, for all values contributing to the integrals in (\ref{eq:pray}).
If $\boldsymbol{\rho}_3$ represents a pure state of linear polarization $\operatorname{diag}\{1,0,0\},$ the ray scattered in the $1$-direction has zero intensity, but such an isolated direction does not affect $P_{\mathrm{Ray}}.$

\section{Comparison with other measures of degree of polarization, and values for sample polarization matrices}\label{sec:exs}

In this section, specific examples of 3-dimensional ensembles represented by particular polarization matrices $\boldsymbol{\rho}_3$ are considered, demonstrating explicitly that $P_{\mathrm{Ray}}$ is not a unitary invariant; progress by means of example is the only way understanding of $P_{\mathrm{Ray}},$ in the absence of a general closed analytic form.
Before discussing these examples, I will describe some other measures which are unitary invariant, that is, they depend solely on the eigenvalues $1 \ge \lambda_1 \ge \lambda_2 \ge \lambda_3 \ge 0.$
The measure of Ref. \onlinecite{edpw:degree}, defined $\lambda_1 - \lambda_2,$ is not a purity measure of polarization under the present definition; however, a slightly adjusted definition is: 
\begin{equation} 
   P_{\mathrm{lin}} = \lambda_1 - \lambda_3,
   \label{eq:wolf}
\end{equation}
which is obviously linear in the eigenvalues.
A second measure, introduced in Ref. \onlinecite{sskf:degree}, depends on a quadratic combination of the eigenvalues, defined
\begin{equation}
   P_{\mathrm{quad}}^2 = [3\tr \boldsymbol{\rho}_3^2 - (\tr\boldsymbol{\rho}_3)^2]/2.
   \label{eq:fri}
\end{equation}
$P_{\mathrm{quad}}^2$ is frequently used in the analysis of quantum mechanical density matrices.\cite{fraassen:quantum}
Another important purity measure for quantum density matrices is the von Neumann entropy,\cite{vonneumann:mathematical} whose analog here is $P_{\mathrm{vN}},$ where
\begin{equation}
   P_{\mathrm{vN}}^2 = 1 + \frac{1}{\log 3}\sum_{j=1}^ 3 \lambda_j \log \lambda_j.
   \label{eq:vne}
\end{equation}
This is the natural purity measure in quantum statistical mechanics,\cite{peres:quantum,vonneumann:mathematical} and is considered here for comparison. 
Each of the measures defined in Eqs. (\ref{eq:wolf}), (\ref{eq:fri}), (\ref{eq:vne}) is a purity measure, and in a neighborhood of the totally unpolarized state, is linear in the eigenvalues.

\begin{figure}
   \centerline{\includegraphics[width=8cm]{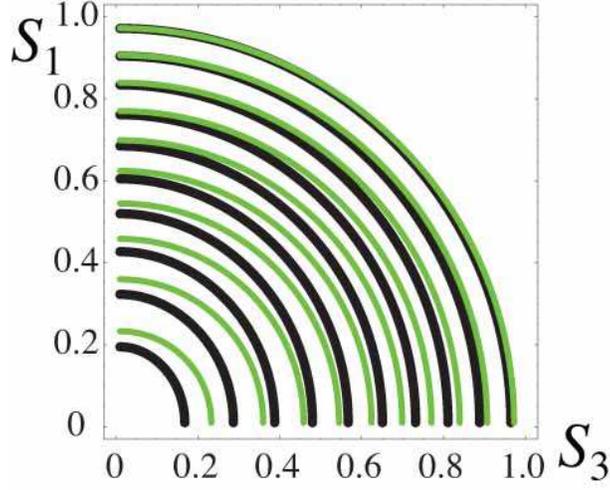}}
   \caption{Contours of constant $P_{\mathrm{Ray}}$ (thick lines) and $P_{\mathrm{quad}}$ (thin green lines) in the $S_1, S_3$-plane, for $3\times 3$ paraxial partial polarization.
   The contour lines of $P_{\mathrm{quad}}$ depend only on the radius $\sqrt{S_1^2 + S_3^2},$ whereas the $P_{\mathrm{Ray}}$ contours have weak angular dependence.
   }
   \label{fig:stokes23}
\end{figure}

\begin{figure}
   \centerline{\includegraphics[width=12cm]{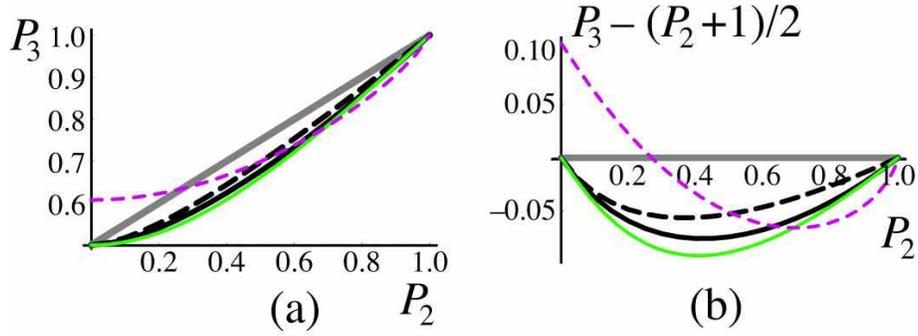}}
   \caption{Illustrating the various 3-dimensional polarization degree measures $P_3$ for paraxial partial polarization, parametrized by the paraxial measure $P_2.$ (a) $P_3$ plot; (b) Plot of $P_3 - (P_2+1)/2$ (i.e. linear part subtracted). 
   $P_3 =$ $P_{\mathrm{Ray}}$ with $S_3$ ($S_1$) $= 0$ (thick (dashed) line); $P_{\mathrm{lin}}$ (thick gray line); $P_{\mathrm{quad}}$ (thin green line), $P_{\mathrm{vN}}$ (thin dashed purple line).
   }
   \label{fig:stokes232}
\end{figure}

An important question is how the various 3-dimensional degrees of polarization measure a paraxially polarized beam; this is the case usually considered in discussions of Rayleigh scattering.
In this case, the choice of cartesian coordinates in Eq. (\ref{eq:rho3def}) eliminates $S_2,$ so $\boldsymbol{\rho}_3$ depends on the two Stokes parameters $S_1$ and $S_3,$ namely
\begin{equation}
   \boldsymbol{\rho}_{3,\mathrm{trans}} = \frac{1}{2}\left(\begin{array}{ccc}
1 + S_1 & - i S_3 & 0 \\
i S_3 & 1 - S_1 & 0 \\
0 & 0 & 0\end{array} \right).
   \label{eq:rho3trans}
\end{equation}
The transverse, two-dimensional degree of polarization is $P_2 = \sqrt{S_1^2+S_3^2}$ (agreeing with the quantity defined in Ref. \onlinecite{edpw:degree}).
However, as mentioned above, the transversally unpolarized state with $S_1 = S_3 = 0,$ has some measure of 3-dimensional polarization, as it appears completely unpolarized only in the propagation direction; its purity can neither be 0 nor 1. 
$P_{\mathrm{Ray}}$ is plotted as a function of $S_1$ and $S_3$ in Fig. \ref{fig:stokes23}; although its main dependence is on $P_2,$ there is weak dependence on the angle $\arctan(S_3/S_1).$
The paraxial partially polarized $\boldsymbol{\rho}_3$ is pure when $P_2 = 1,$ and all definitions of polarization agree here. 
When $P_2 = 0,$ $P_{\mathrm{Ray}} = P_{\mathrm{quad}} = P_{\mathrm{lin}} = 1/2,$ and $P_{\mathrm{vN}} = \sqrt{1-\log2 / \log 3} \approx 0.608.$
The 3-dimensional degrees of polarization for this example are plotted in Fig. \ref{fig:stokes232}.

Two different 3-dimensional polarization matrices with the same eigenvalues were considered in Ref. \onlinecite{dennis:geometric}.
Generalizations of those examples are as follows.
The first is the diagonal matrix $\boldsymbol{\rho}_{3} = \diag\{1+a,1,1-a\}/3,$ with $0 \le a \le 1;$ this is completely unpolarized when $a = 0.$
It is represented geometrically by an ellipsoid whose axes are the diagonal elements, and the mean angular momentum vector $\boldsymbol{N}$ is zero.
The second example is defined such that each $M_j = 1/3$ (the ellipsoid is a sphere) and $|\boldsymbol{N}| = N_3 = a/3$ ($0 \le a \le 1$).
For example 2, $I(\theta, \phi)$ is uniform, and $P_2(\theta,\phi)/I(\theta,\phi) = a | \cos\theta|.$
The two examples have the same eigenvalues $(1+a)/3, 1/3, (1-a)/3,$ but geometrically are very different.
In both cases, $P_{\mathrm{Ray}}$ depends linearly on $a,$ with 
\begin{equation}
   P_{\mathrm{Ray,ex1}} = 0.5932 a, \; P_{\mathrm{Ray,ex2}} = 0.5 a,
   \label{eq:praxex12}
\end{equation}
where the gradient for $P_{\mathrm{Ray,ex1}}$ is determined by numerical integration.
The two ensembles thus have different Rayleigh-defined degrees of polarization for all $a > 0.$
By comparison, unitary invariance means the three other measures do not distinguish between the two ensembles, with $P_{\mathrm{lin}} = 2a/3,$ and $P_{\mathrm{quad}} = a/\sqrt{3}$ ($P_{\mathrm{vN}}$ is approximately linear in $a,$ but cannot be expressed simply).

The final example considered here is a geometrically simple case in which the polarization matrix is real and degenerate (its ellipsoid is axisymmetric and $\boldsymbol{N} = 0$), i.e. $\boldsymbol{\rho}_{3} = \mathbf{M} = \diag\{2m, 1-m, 1-m\}/2$ for $0 \le m \le 1.$
In this case, $P_{\mathrm{Ray}}$ can be found analytically, and $P_{\mathrm{Ray}} = P_{\mathrm{lin}} = P_{\mathrm{quad}} = | 3m - 1|/2,$ and $P_{\mathrm{vN}}$ is numerically close.

\section{Discussion}\label{sec:dis}

Elements of the polarization matrices $\boldsymbol{\rho}_2, \boldsymbol{\rho}_3$ are complex when the polarization states in the underlying ensembles have some elliptical or circular polarization; in $\boldsymbol{\rho}_2,$ this is given by a nonzero value of the third Stokes parameter $S_3,$ and in $\boldsymbol{\rho}_3,$ by a nonzero net angular momentum vector $\boldsymbol{N}.$
Therefore, unitary invariance implies that states of elliptic polarization are treated equivalently to states of linear polarization.
This is clearly appropriate for paraxially propagating states of polarization, where all polarization states are on the same footing on the Poincar{\'e} sphere, and may be freely transformed to each other by retarders and rotators without changing the degree of polarization.

The situation is less clear in nonparaxial physics, where there are several well-known deviations from the paraxial case, such as the polarization topology of inhomogeneous 3-dimensional fields\cite{nh:wavestructure,bd:324}, and the geometric phase in twisted optical fibers.\cite{cw:manifestations,berry:159}
This difference originates in the fact that the 3-dimensional orientation of a state of linear polarization requires two parameters (the direction of the real $\boldsymbol{E}$ vector), but the orientation of elliptic polarization requires three (the Euler angles of the polarization ellipse).
In Rayleigh scattering, a ray's polarization is given by the projection of the 3-dimensional polarization matrix into the transverse plane of the ray by Eq. (\ref{eq:proj}); complex states of elliptic polarization project over all directions differently than real states of linear polarization, and this is manifest in the lack of unitary invariance in $P_{\mathrm{Ray}},$ ultimately due to the non-unitarity of the physical mechanism of Rayleigh scattering.
Given the geometric difference between nonparaxial and paraxial polarization, it seems that appealing to analogy with the two-dimensional paraxial case is insufficient motivation by itself for demanding the unitary invariance of the 3-dimensional degree of polarization; consideration of other physical instances of the 3-dimensional polarization matrix, such as in near-field diffraction, or light in a twisted fiber, may lead to other physically-motivated definitions of the polarization matrix that are, or are not, unitarily invariant.

The definition (\ref{eq:pray}) can be modified mathematically to give a unitarily invariant version of $P_{\mathrm{Ray}},$ at the cost of the physical interpretation.
Instead of averaging over symmetric projections perpendicular to the observation direction $\boldsymbol{u}(\theta,\phi),$ one integrates over complex hermitian projections that are orthogonal to all complex vectors (i.e. all polarization states).
This average respects unitary invariance, and $\boldsymbol{\rho}_3$ can be diagonalized before the average is taken.
Therefore, under this unitary invariant measure, the partially polarized paraxial ensembles behave like $P_{\mathrm{Ray}}$ with $S_3 = 0$ (plotted in Fig. \ref{fig:stokes232}), the examples 1 and 2 both behave like example 1, and any degenerate $\boldsymbol{\rho}_3$ as the axisymmetric example.

The geometrical interpretation of the 3-dimensional polarization matrix was used in Ref. \onlinecite{hannay:polarization} to describe the polarization of skylight using a canopy atmosphere model.
Incident sunlight is Rayleigh scattered in a thin layer of fixed height above the plane of the observer, giving rise to $\boldsymbol{\rho}_3$ independent of position; the polarization in an observation direction was found using an orthogonal projection of this matrix (by Eq. (\ref{eq:proj})); in particular, the neutral points,\cite{chandrasekhar:radiative,bdl:373} where the degree of polarization vanishes, correspond to the Maxwell axes of the ellipsoid.\cite{dennis:canonical}

Clearly, the definition of $P_{\mathrm{Ray}}$ in (\ref{eq:pray}) is not the only way $P_2(\theta,\phi)$ and $I(\theta,\phi)$ can be combined to give a well-defined degree of polarization; for instance, any positive powers can be used, and $I(\theta,\phi)$ need not be included.
In particular, the choice $P_2(\theta,\phi)^2 I(\theta,\phi)^2$ in the integrand can be easily integrated, yielding $2 [3\tr \mathbf{M}^2 - (\tr\mathbf{M})^2 + 5\boldsymbol{N}\cdot\boldsymbol{N}]/[\tr\mathbf{M}^2 + 3 (\tr\mathbf{M})^2].$
The choice of (\ref{eq:pray}) is made here not only because its definition is physically natural, but also it is closer, at least in the examples considered, to the measures $P_{\mathrm{lin}}$ and $P_{\mathrm{quad}}.$ 
In quantum mechanics, it is more usual to consider the squares of the purity measures considered in optics (such as the measures in Eqs. (\ref{eq:fri}) and (\ref{eq:vne})); the squares are more natural from the point of view of mathematical definition, but do not appear directly (in two dimensions) in terms of the elegant decomposition $\boldsymbol{\rho}_2 = P_2 \boldsymbol{\rho}_{\mathrm{pure}} + P_2 \boldsymbol{\rho}_{\mathrm{2,un}}.$

It may be possible to determine the Rayleigh degree of polarization directly by experimental measurement.
A possible Rayleigh scatterer in such an experiment would be a metallic nanoparticle, embedded in some transparent medium; this would be placed in an appropriate incoherent light field represented by the desired $\boldsymbol{\rho}_3,$ sufficiently constant over the lengthscale of the particle.
It would only be necessary to measure the polarization of $\boldsymbol{\rho}_2(\theta,\phi)$ over a hemisphere, as the projections in Eq. (\ref{eq:proj}) do not distinguish $\pm \boldsymbol{u}(\theta,\phi)$ -- measuring the polarization over a large number of scattering directions may cause technical problems, and further discussion is outside the scope of the present work.
Of course, the average over all directions of the scattered ray polarization depends on the kind of scatterer (anisotropic Rayleigh, Mie, etc); it is likely that different, scatterer-dependent measures of the degree of polarization can be defined.
However, these other scatterers have some internal structure, unlike an isotropic Rayleigh scatterer, and the Rayleigh degree of polarization depends only on the electromagnetic polarization matrix $\boldsymbol{\rho}_3.$

Projecting a density matrix on subspaces is reminiscent of Gleason's theorem\cite{fraassen:quantum,gleason:measures} in quantum mechanics, in which projection assigns weights to 1-dimensional subspaces.
However, averaging such one-dimensional projections (over all hermitian projections, being quantum mechanical) only gives the trace; this construction completely misses the complicated structure of $P_{\mathrm{Ray}},$ and even its unitary counterpart.

The 3-dimensional polarization matrix is insufficient to describe propagation of partially coherent nonparaxial fields, since propagation direction information is also required (in the form, for instance, of a Wigner function\cite{alonso:wigner}).
It is not clear physically how a 3-dimensional degree of polarization ought to behave under propagation, especially if it is assumed to depend only on the polarization matrix.
Mechanisms such as dipole scattering, however, are completely independent of the propagation directions of the electromagnetic fields giving rise to the polarization matrix (neglecting recoil effects), and are therefore a natural place to find physical relevance of measures of the 3-dimensional degree of polarization.
However, as has been shown, the natural measure based on this process does not depend only on the eigenvalues of $\boldsymbol{\rho}_3$ (it is not a unitary invariant), unlike other measures which have been proposed.
The assumptions behind the definition of the 3-dimensional degree of polarization therefore require further consideration.

\section*{Acknowledgements} 
I am grateful to Miguel Alonso, Jeremy Baumberg, Giampaolo D'Alessandro, John Hannay, and Emil Wolf for discussions.
My research is supported by the Royal Society of London.

%\bibliographystyle{osajnl}
%\bibliography{mycites}

\end{document}